\documentclass[format=sigconf]{acmart}

\theoremstyle{definition}

\usepackage{mathrsfs}
\usepackage{amsmath}
\usepackage{makecell}
\usepackage{booktabs}
\usepackage{dblfloatfix}
\usepackage{fixltx2e}
\usepackage{multirow}

\usepackage[algo2e,ruled,vlined]{algorithm2e}
\SetCommentSty{scriptsize}
\SetKwComment{tcc}{\{}{\}}
\SetKwInput{Input}{Input}
\SetKwInput{Output}{return}
\SetKwIF{If}{ElseIf}{Else}{if}{}{else if}{else}{endif}
\SetKwFor{While}{while}{}{endw}
\SetKwFor{ForEach}{for each}{}{endfch}
\newcommand{\code}[1]{{\textnormal{\texttt{#1}}}}

\graphicspath{{./figures/}}

\author{Amanda Bienz}
\affiliation{University of Illinois at Urbana-Champaign}
\email{bienz2@illinois.edu}
\author{William D. Gropp}
\affiliation{University of Illinois at Urbana-Champaign}
\email{wgropp@illinois.edu}
\author{Luke N. Olson}
\affiliation{University of Illinois at Urbana-Champaign}
\email{lukeo@illinois.edu}

\title{Improving Performance Models for Irregular Point-to-Point Communication}
\begin{abstract}
    Parallel applications are often unable to take full advantage of emerging
    parallel architectures due to scaling limitations, which arise due to
    inter-process communication.  Performance models are used to analyze the
    sources of communication costs.  However, traditional models for
    point-to-point communication fail to capture the full cost of many irregular
    operations, such as sparse matrix methods.  In this paper, a node-aware
    based model is presented.
    Furthermore, the model is
    extended to include communication queue search time as well as an
    additional parameter estimating network contention.  The resulting
    model is applied to a variety of irregular communication patterns throughout
    matrix operations, displaying improved accuracy over traditional models.
\end{abstract}

\acmConference{EuroMPI 2018}{September 23--26, 2018}{Barcelona, Spain}

\begin{document}
\maketitle

\section{Introduction}
As parallel computers advance, improvements to hardware yield potential for
solving increasingly large and difficult problems.  However, applications are
often unable to take full advantage of state-of-the-art architectures due to
scaling limitations, which result from inter-process communication costs.  The
cost associated with communication depends on a large number of factors, and
varies across parallel systems, specific partitions, and application scale.
Therefore, performance models are used to analyze the sources of communication
costs among different architectures and network partitions.  Accurate
performance models specify whether the cost is due mainly to the number of
messages communicated, number of bytes transported, distance of transported
bytes, or some other factor.

Traditional models estimate point-to-point communication as a combination
message latency and the cost of transporting bytes.  Irregular operations, such
as sparse matrix methods, acquire costs that are not captured by traditional
models.  Figure~\ref{figure:tradition_spgemm_model} displays the measured and
modeled communication costs acquired when performing a sparse matrix-matrix
(SpGEMM) multiply on the levels of an algebraic multigrid (AMG) hierarchy for an
unstructured linear elasticity matrix.
\begin{figure}[ht!]
    \centering
    \includegraphics[width=0.45\textwidth]{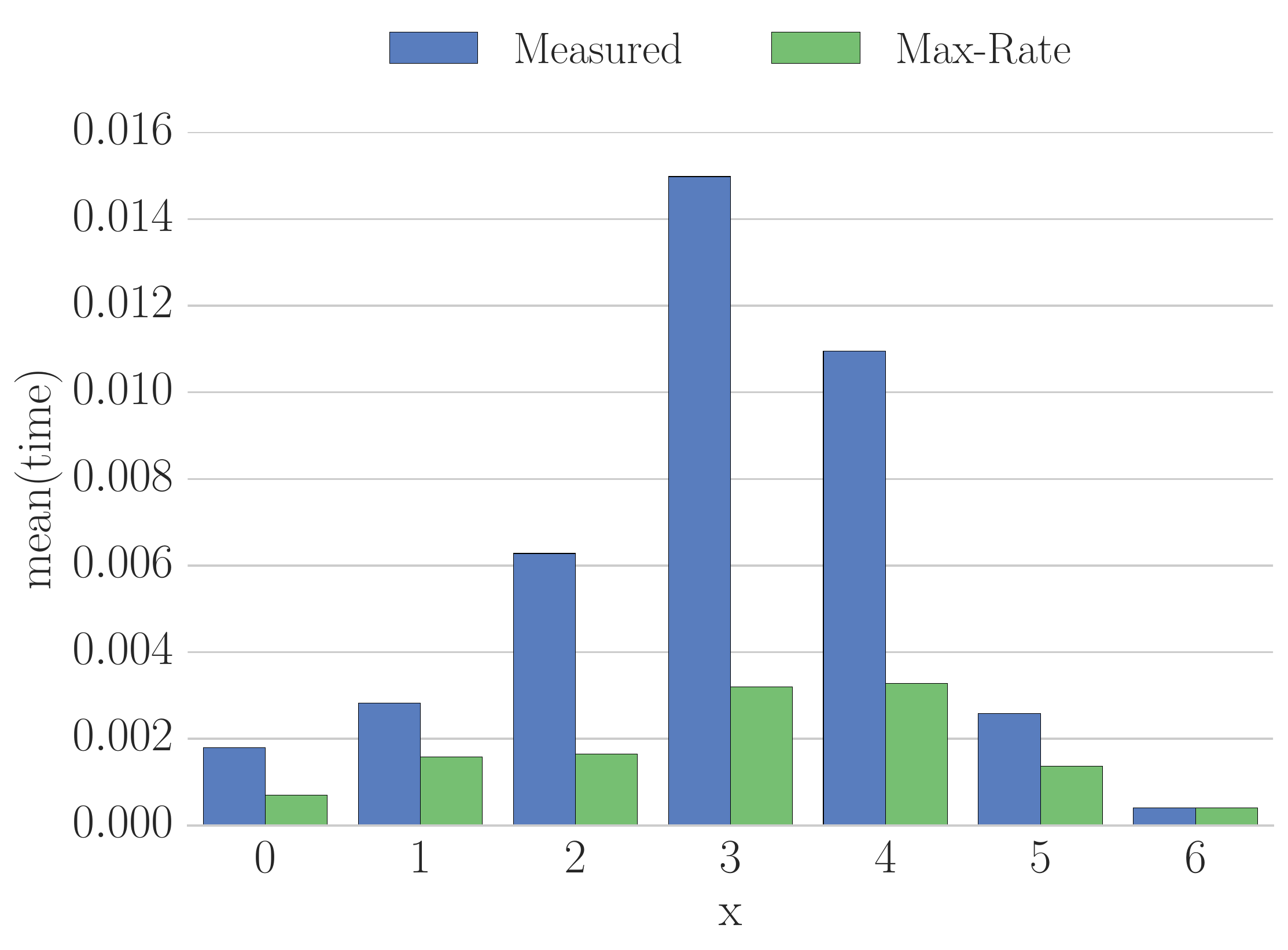}
    \caption{Measured and modeled communication costs associated with a SpGEMM
    on each level of a linear elasticity AMG hierarchy, on $8\,192$ processes of
    Blue Waters supercomputer.}\label{figure:traditional_spgemm_model}
\end{figure}
These timings, as well as the associated model parameters, are for $8\,192$
processes of Blue Waters supercomputer~\footnote{\url{https://bluewaters.ncsa.illinois.edu/}}~\cite{BlueWaters13}.  The
traditional postal model results in nearly identical timings to more robust
models, such as the max-rate model~\cite{MaxRate}, which takes into account the limitations of multiple
communicating processes per node.  For this problem both models capture only a
fraction of the measured time.

This paper extends traditional performance models to accurately model the
irregular point-to-point communication that occurs in commonly used operations.
This paper presents three novel contributions, including an improvement to the
max-rate model~\cite{MaxRate} measurements:
\begin{enumerate}
  \item node-aware model parameters;
  \item an extension to include quadratic queue search times in communication; and
  \item an additional parameter estimating network contention.
\end{enumerate}
The remainder of this paper
is outlined as follows.  Section~\ref{section:background} describes parallel
point-to-point communication as well as corresponding traditional performance
models.  Node-awareness is added to traditional models in
Section~\ref{section:node_aware}.  Section~\ref{section:models} describes a
high-volume ping-pong algorithm along with additional acquired penalties, with a
queue search parameter described in Section~\ref{section:queue_search} and a
network contention penalty in Section~\ref{section:network_contention}.  The
improved model parameters are applied to commonly used operations and compared
against measured timings in Section~\ref{section:applications}.  Finally,
conclusions, limitations, and future directions are described in
Section~\ref{section:conclusion}.

\section{Background}\label{section:background}

Many common parallel operations, such as those involving sparse matrices,
require MPI point-to-point communication.  This category of communication
consists of sending a single message between a set of processes.  In a typical
implementation, the pairs of communicating processes, along with the size of
associated messages, vary. The point-to-point communication procedure varies
with MPI implementation.  Typically, each message consists of both an envelope
and data, where the envelope contains a message description including the tag,
MPI communicator, message length, and process of origin.  There are a variety of
methods for sending data, such as sending the data immediately or waiting for
the receive process to allocate buffer space.  In the implementations
investigated in this paper, a message is communicated via a specific protocol of
short, eager, or rendezvous, based on message size.  The short protocol consists of
sending very small messages as part of the envelope directly between processes.
Messages that are too large to fit in the envelope, but remain relatively small,
are communicated with eager protocol.  This protocol assumes buffer space is
available, and immediately communicates the data to the receiving process.
Lastly, sufficiently large messages are communicated with rendezvous protocol,
during which the envelope is communicated first, and the remainder of the data
is only communicated after the receiving process has allocated buffer space.

The cost associated with each message is dependent on the time required to
initialize communication as well as the per-byte transport cost.  Therefore,
short protocol is significantly less costly than the others as only a single
envelope is communicated, yielding minimal costs associated with both latency
and bandwidth.  Messages communicated with eager protocol have low latency costs
as the messages are sent directly between processes.  However, the associated
per-byte transport cost increases, as these messages can require significant
amounts of buffering.  Finally, rendezvous messages yield low per-byte transport
costs but increased latency requirements associated with initial envelope
communication and synchronization.

The traditional postal model estimates the cost of communicating a message as
the sum of the message startup cost and the per-byte transport, with separate
parameters for each message protocol.  This can be defined as
\begin{equation}\label{equation:postal}
    T = \alpha + \beta \cdot s,
\end{equation}
where $\alpha$ is the latency, $\beta$ is the cost to transport a byte of data,
and $s$ is the number of bytes to be transported.  As the associated costs vary
with message protocol, separate values for $\alpha$ and $\beta$ are used when
communicating with short, eager, and rendezvous protocols.  This model
accurately analyzes the cost of a standard ping-pong test, in which two
processes are sending messages to one another.  However, it fails to account for
a variety of penalties that occur during communication in typical operations on
state-of-the-art supercomputers.

There are many alternatives to the postal model that account for many penalties
that arise in standard supercomputers.  The LogP model splits the $\alpha$
into latency, the cost required by the hardware, and overhead, the cost
associated with the software~\cite{LogP, PRAM}.  This addition of overhead
allows this model to capture the cost of overlapping communication and
computation.  The LogGP model extends the LogP model to analyze the cost of long
messages~\cite{LogGP}.  Network contention parameters are investigated with the
LoPC and LoGPC models~\cite{LoPC, LoGPC}.  Accurate models exist for network
contention in collective communication, such as the $\code{MPI\_Alltoall}$
operation~\cite{ContentionAlltoall}.  Futhermore, learning algorithms yield
accurate contention prediction~\cite{PredictContention}.  Topology-awareness can
improve models, as hop count, or the number of links traversed by a message,
affects the cost of communication~\cite{HopModel}.  Computer simulations can
accurately estimate the cost of communication, but at a significantly increased
cost~\cite{Bigsim, ORCS, LogGOPSim}.  Network contention has been previously
modeled for collective communication, specifically the $\code{MPI\_Alltoall}$
operation.

The max-rate model~\cite{MaxRate} improves upon the postal model by defining the cost of
communication as dependent on not only the latency and inter-process bandwidth
costs, but also on the maximum bandwidth by which a node can inject data into
the network.  Therefore, this model accounts for the fact that injection
bandwidth becomes a bottleneck when communicating from four or more processes
per node, as is typical with state-of-the-art parallel computers~\cite{MaxRate}.

Throughout the remainder of this paper, the max-rate model is used as a
baseline.  All ping pong timings are collected through multiple runs of
Baseenv~\footnote{\url{http://wgropp.cs.illinois.edu/projects/software/index.html}}, a topology-aware library useful for benchmarking
performance.  Each ping-pong test consists of four duplicate timings, with
exception to the original max-rate tests, which test the various numbers of
actively communicating processes-per-node one time each.  Each Baseenv program
is tested three different times.

This models throughout this paper are tested with Blue Waters, a Cray XE/XK machine at the
National Center for Supercomputing Applications (NCSA) at University of
Illinois.  Blue Waters contains a 3D torus Gemini interconnect, in which each Gemini consists of
two nodes.  The system contains $22\,636$ XE compute nodes, each comprised of
two AMD 6276 Interlagos processors, as well as $4\,228$ XK compute nodes
containing a single AMG processor along with an NVIDIA GK110 Kepler GPU\@.  All
tests in this paper are performed on partitions of XE system nodes.  The tests
use a CrayMPI implementation that is similar to MPICH\@.

\section{Node-Aware Modeling}\label{section:node_aware}
\begin{figure}[ht!]
    \centering
    \includegraphics[width=0.5\textwidth]{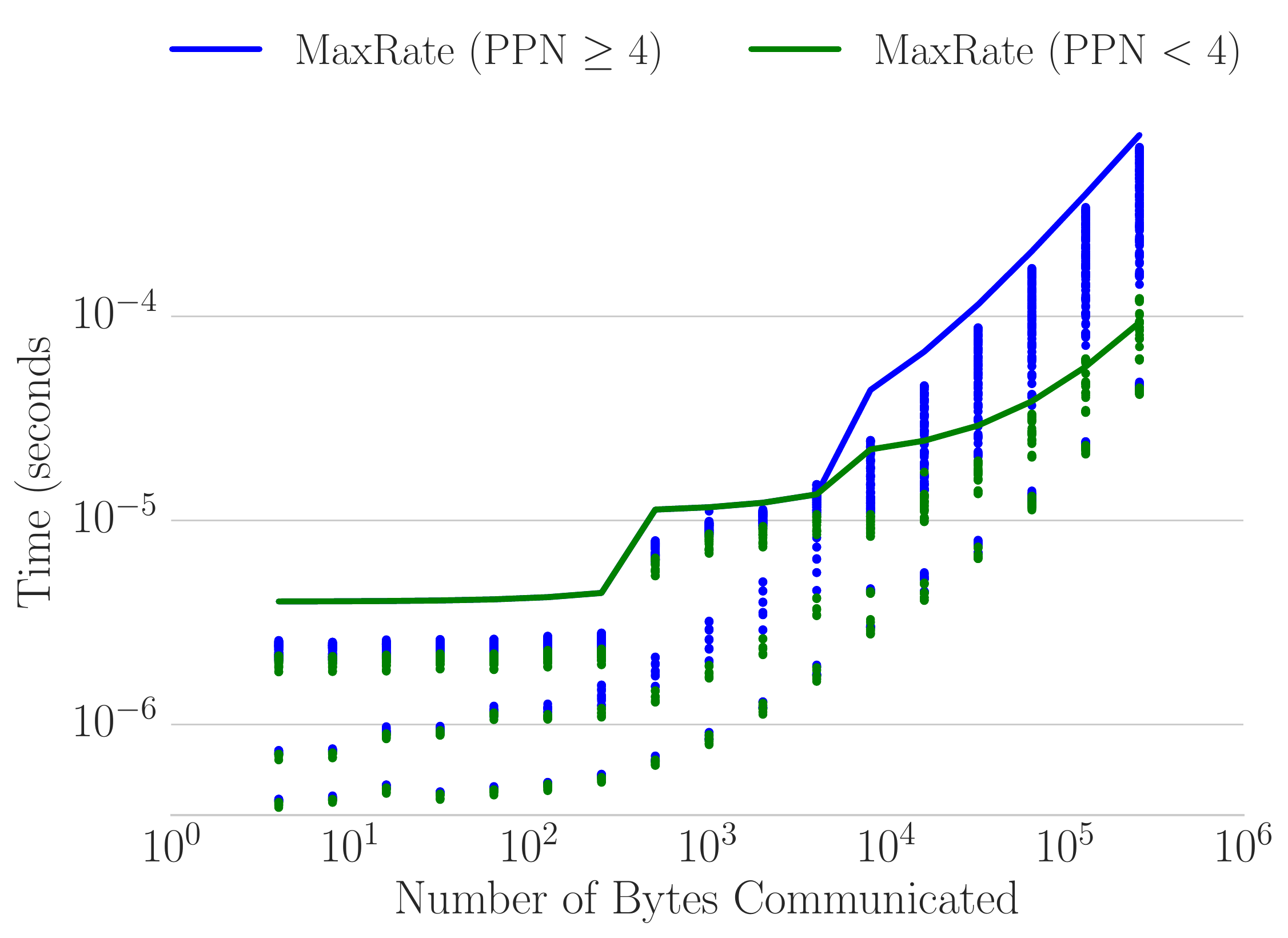}
    \caption{Ping-pong measured times (dots) versus max-rate model (lines) on
    Blue Waters using parameters from~\cite{TODO}.}\label{figure:max_rate_standard}
\end{figure}

The max-rate model is defined as
\begin{equation}
    T = \alpha + \frac{\texttt{ppn} \cdot s}{\min(R_{N}, \texttt{ppn} \cdot R_{b})},
    \label{equation:max_rate}
\end{equation}
where $\texttt{ppn}$ is the number of actively communicating processes per node,
$R_{b}$ is the rate at which data can be sent between two processes, or the
inverse of $\beta$, and $R_{N}$ is the maximum rate at which a node can inject
data into the network.  Therefore, when the value of $\texttt{ppn} \cdot R_{b}$
is less than injection bandwidth, this model reduces to the postal model.
However, with a sufficiently large number of active processes per node, the
per-byte transport rate is measured as injection bandwidth.
Figure~\ref{figure:max_rate_standard} displays the max-rate model versus
measured times when communicating a single message of various size between pairs
of processes.
These associated models are computed with published Blue Waters
parameters~\cite{MaxRate}, and the measured times are acquired from sending
messages between processes that lie on the same socket, different sockets of the
same node, or on neighboring nodes of Blue Waters.  While the addition of
network injection limits yields large improvement over the standard postal model
when communicating rendezvous messages from a large number of processes per
node, the model overestimates for a large portion of these timings.

\begin{figure}[ht!]
    \centering
    \includegraphics[width=0.45\textwidth]{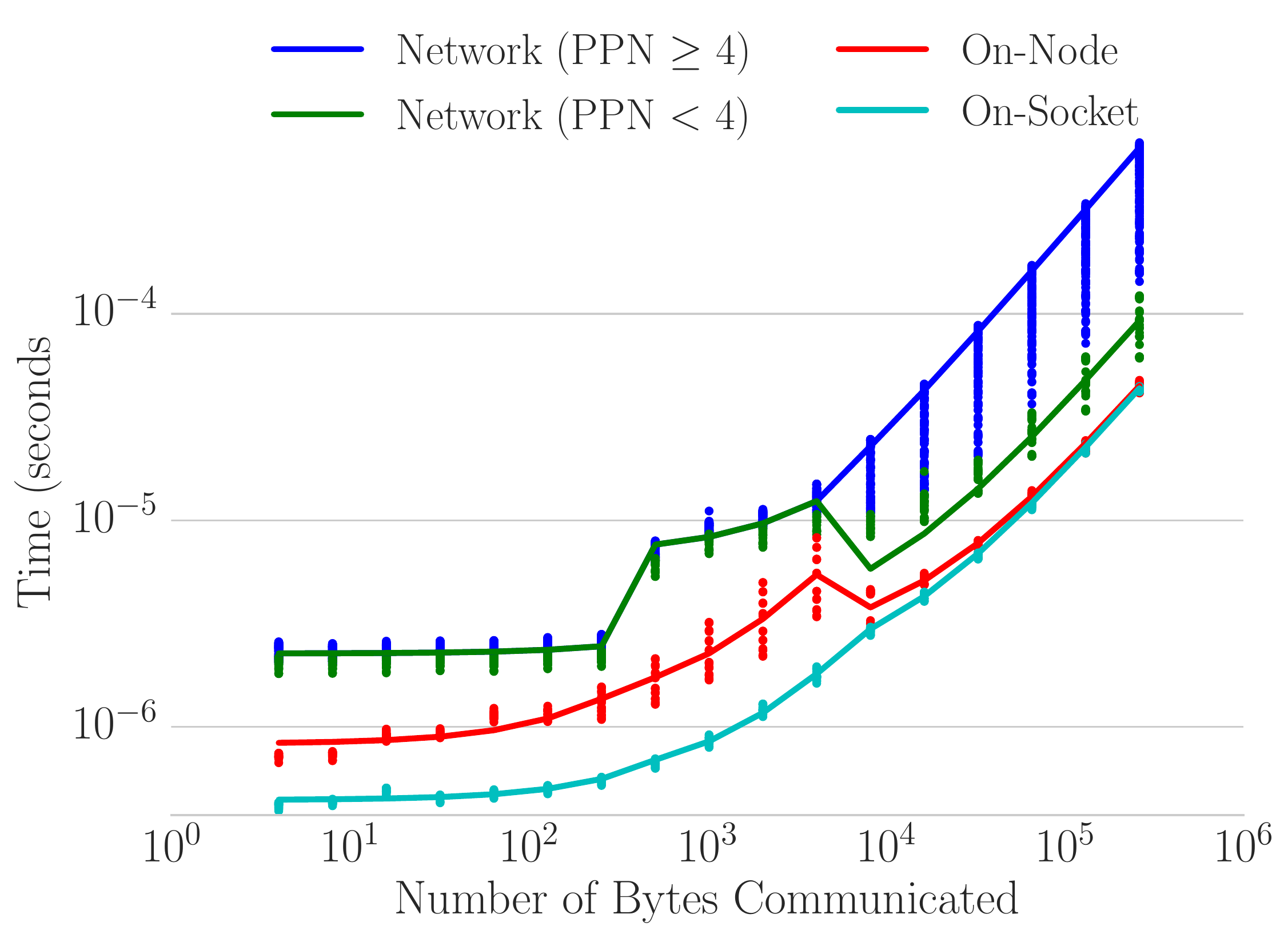}
    \caption{The max-rate model versus measured times when split into on-socket,
    on-node but off-socket, and off-node communication.}\label{figure:max_rate_split}
\end{figure}

There is a large difference between intra-socket messages, intra-node messages
that traverse across sockets, and communication between two nodes.  Therefore,
different parameters should be used for each of these cases.  Furthermore,
intra-node messages are not injected into the network, and therefore the simple
postal model is sufficient.  Figure~\ref{figure:max_rate_split} displays the
measured versus modeled times after splitting the model into on-socket, on-node
but off-socket, and off-node messages for both Blue Waters.
The parameters corresponding this node-aware model are listed in
Table~\ref{table:max_rate_split}.
\begin{table*}[t]
    \centering
    \begin{tabular}{l c c c c c c c }
      \toprule
         & \multicolumn{2}{c}{intra-socket} & \multicolumn{2}{c}{intra-node} &
        \multicolumn{3}{c}{inter-node} \\
         & $\alpha$ & $R_{b}$ & $\alpha$ & $R_{b}$ & $\alpha$ & $R_{b}$ & $R_{N}$\\
        \midrule
          \multicolumn{1}{l}{short} & 4.4e-07 & 2.2e09 & 8.3e-07 & 4.8e08
            & 2.3e-06 & 1.3e09 & $\infty$ \\
          \multicolumn{1}{l}{eager} & 5.3e-07 & 3.2e09 & 1.2e-06 & 9.6e08
            & 7.0e-06 & 7.5e08 & $\infty$ \\
          \multicolumn{1}{l}{rend} & 1.7e-06 & 6.2e09 & 2.5e-06 & 6.2e09
            & 3.0e-06 & 2.9e09 & 6.6e09\\
        \bottomrule
    \end{tabular}
    \caption{Parameters for node-aware max-rate model on Blue Waters.}\label{table:max_rate_split}
\end{table*}

\section{Additional Penalties}\label{section:models}

Realistic applications that involve point-to-point communication typically
require more than one message to be communicated from any process.  However,
standard communication models were created to analyze the cost of sending a
single message, and do not extend accurately to large message counts.  A
ping-pong test with large message counts, described in
Algorithm~\ref{algorithm:ping_pong}, acquires additional costs not captured by
the postal or max-rate models. \\

\begin{algorithm2e}[ht!]
  \DontPrintSemicolon%
	\KwIn{%
    \begin{tabular}[t]{l l}
        $\texttt{rank}$: & MPI rank of current process\\
        $p$: & process with which to communicate\\
        $n$: & number of messages to communicate\\
        $s$: & size of each message (in bytes)\\
        $\texttt{send\_tags}$: & list of $n$  MPI send tags\\
        $\texttt{recv\_tags}$: & list of $n$ MPI receive tags\\
    \end{tabular}
  }
  \;
    \eIf{$\texttt{rank} < p$}{%
        \For{$i < n$}{%
            MPI\_Isend($\dots$, $s$, $\dots$, $p$, $\texttt{send\_tags}_{i}$, $\dots$)\;
        }
        MPI\_Waitall($n$, $\dots$)\;
        \;
        \For{$i < n$}{%
            MPI\_Irecv($\dots$, $s$, $\dots$, $p$, $\texttt{recv\_tags}_{i}$, $\dots$)\;
        }
        MPI\_Waitall($n$, $\dots$)\;
    }
    {%
        \For{$i < n$}{%
            MPI\_Irecv($\dots$, $s$, $\dots$, $p$, $\texttt{recv\_tags}_{i}$, $\dots$)\;
        }
        MPI\_Waitall($n$, $\dots$)\;
        \;
        \For{$i < n$}{%
            MPI\_Isend($\dots$, $s$, $\dots$, $p$, $\texttt{send\_tags}_{i}$, $\dots$)\;
        }
        MPI\_Waitall($n$, $\dots$)\;
    }\;

   	\caption{\code{HighVolumePingPong}}\label{algorithm:ping_pong}
\end{algorithm2e}

\subsection{Queue Search}\label{section:queue_search}

Point-to-point communication conceptually requires multiple queues to be formed
and traversed, including a send queue, comprised of sends that have been posted;
a receive queue of similarly posted receives; and an unexpected message queue,
containing messages that have been communicated for which no matching receive
has been posted~\cite{MPIQueues}.  The function and availability of these
queues, which are dependent on MPI implementation, greatly affects the
performance of communicating a large number of messages.

The standard implementation of MPICH creates two separate receive queues, one
for the posted messages and the other for unexpected
messages~\cite{QueueSearch}.  When an envelope is received, the queue of posted
messages is searched for a message in which all variables such as tag, datatype,
communicator, and sending process match the envelope.  If no associated message
has been posted, the envelope and any corresponding data is added to the
unexpected message queue.  Similarly, when a message is posted by the
application, the unexpected message queue is traversed for any corresponding
envelope.  If no match is found, the message is added to the posted message
queue.

The CrayMPI implementation requires a receive queue to be
searched linearly, yielding an additional cost when communicating multiple
messages.  In the worst case, the messages are received in the order opposite of
which they are posted, requiring a traversal of an entire queue for each
receive.  Therefore, the queue search is an $\mathcal{O}(n^{2})$ operation.
Methods have been created to reduce this queue search cost, such as the use of
multiple queues in combination with hash maps~\cite{MultiQueueSearch}.  However,
as a standard queue search is currently implemented in the version of MPI on
Blue Waters, the large queue search cost is investigated.

\begin{figure*}[t]
    \centering
    \includegraphics[width=0.45\textwidth]{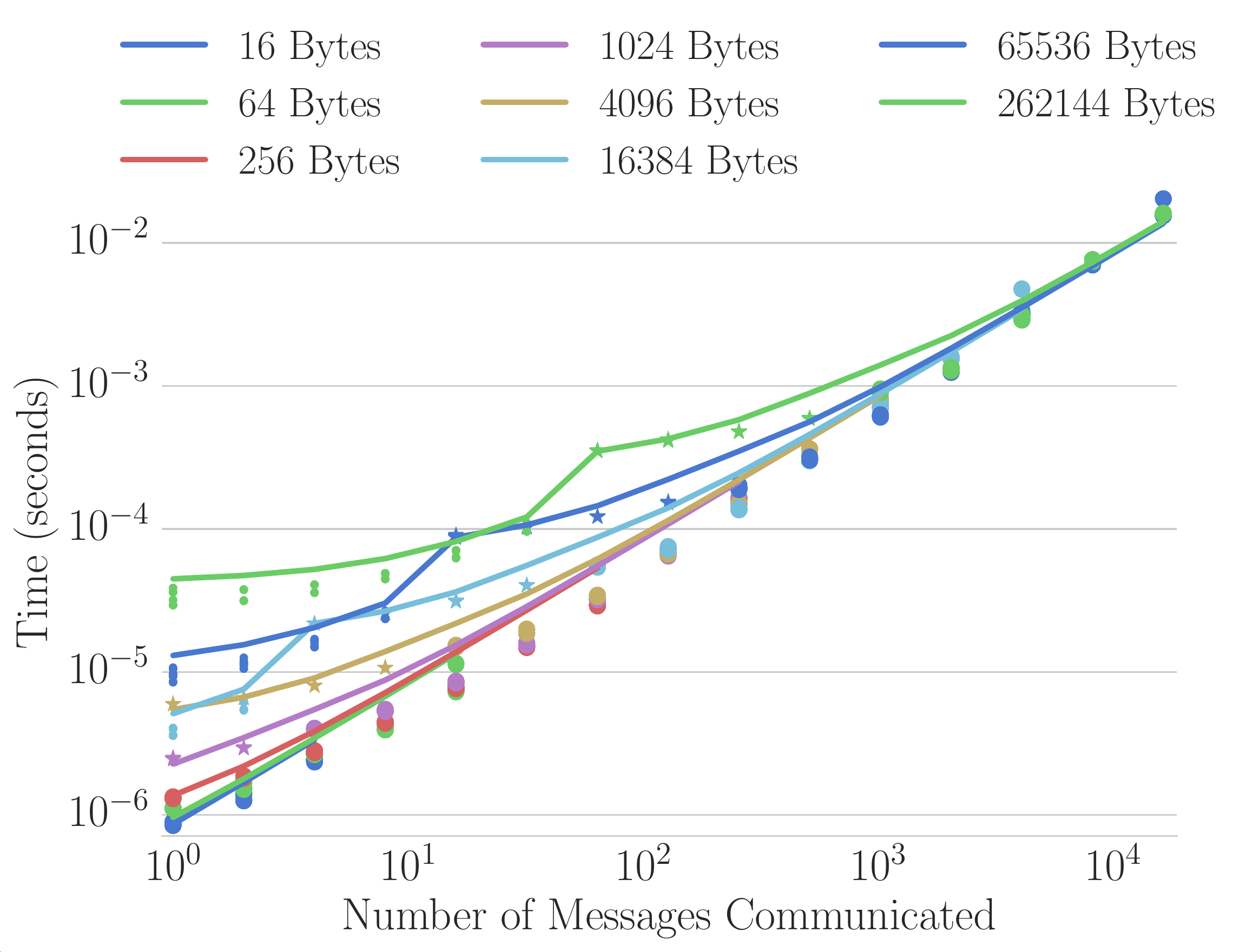}
    \includegraphics[width=0.45\textwidth]{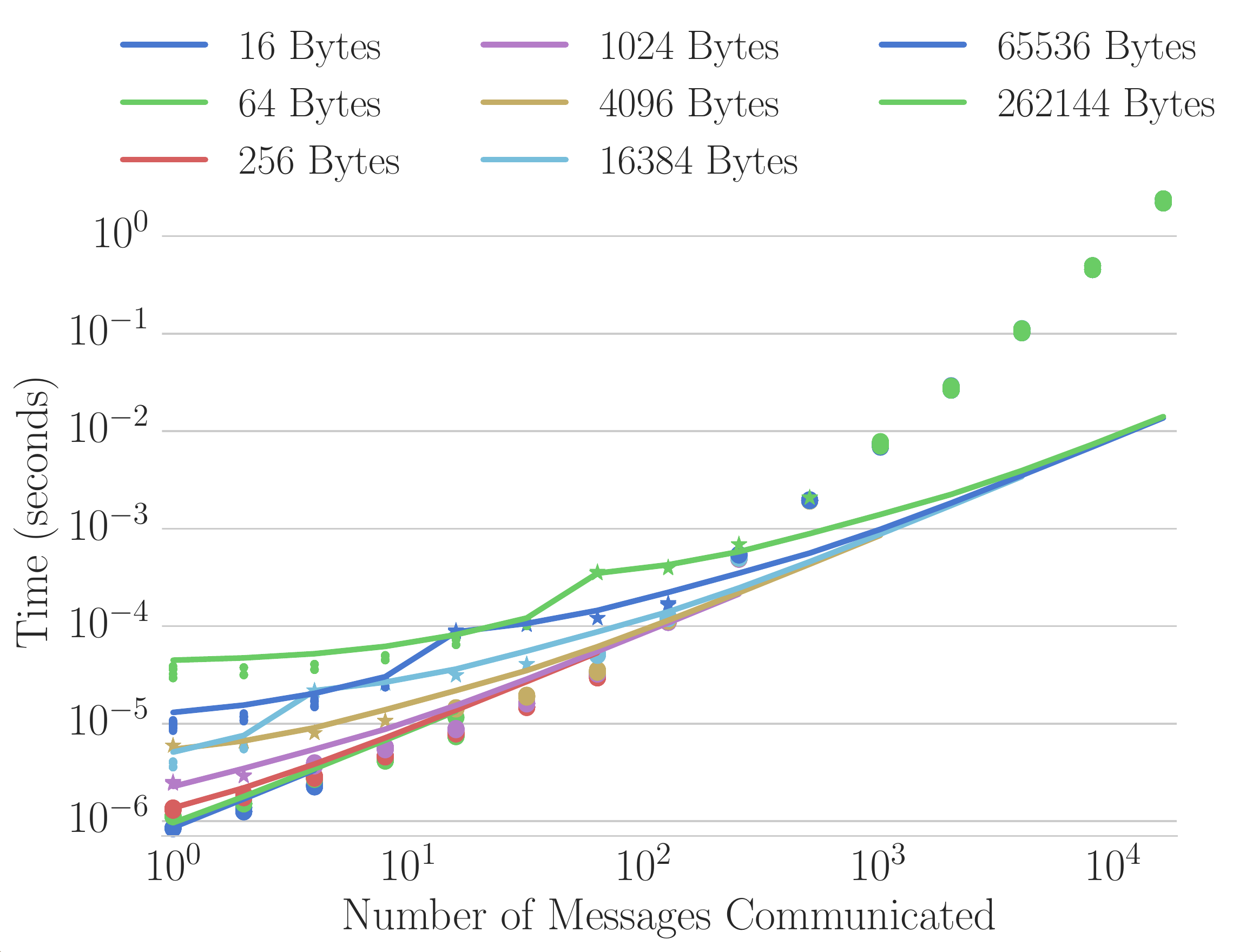}
    \caption{The measured versus modeled (max-rate) costs of sending a number of
    messages between two processes that lie on the same node of Blue Waters.  On
    the left, all receives are posted in the same order that messages are
    received, resulting in no queue search cost.  The right plot displays the
    cost when receives are posted in the opposite order from which they are
    received, resulted in a quadratic queue search cost that is not captured in
    the max-rate model.}\label{figure:queue_not_modeled}
\end{figure*}

Figure~\ref{figure:queue_not_modeled} displays both the measured and modeled
costs for performing the \texttt{HighVolumePingPong} described in
Algorithm~\ref{algorithm:ping_pong} among all $16$ processes local to a single
node on Blue Waters.
The number of messages communicated ranges from $1$ to $10\,000$ with the total
number of bytes injected into the network remaining constant.  In the ideal
scenario, the variables $\texttt{send\_tags}_{i}$ is equal to
$\texttt{recv\_tags}_{i}$ for all $i < n$, resulting in messages being
received in the same order as they are posted.  Therefore, the first message in
the searched queue yields a match, resulting in an $\mathcal{O}(n)$ queue search
cost.  As a result, the max-rate model accurately analyzes these measured times.
In the worst-cast scenario, the variables $\texttt{send\_tags}_{i}$ is equal to
$\texttt{recv\_tags}_{n-i-1}$ for all $i < n$, posting receives in the opposite
order from which messages are received.  Therefore, the entire queue is
traversed for each receive, resulting in an $\mathcal{O}(n^{2})$ queue search
cost, yielding measured times that vary greatly from the model.

The large inaccuracies of traditional models for large message counts motivates
adding an additional parameter for the time required to search the receive
queues.  This addition to the models is defined as
\begin{equation}
    T_{q} = \gamma \cdot n^{2},
\end{equation}
where $\gamma$ is the cost of stepping through either the posted or unexpected
message queue.  This cost is independent of message sending protocol as well as
relative locations of the send and receive processes.  Therefore, there is a
single parameter for all combinations of on-socket, on-node, off-node, and
short, eager, and rendezvous.  The upper bound queue search cost is described as
\begin{equation}
    \gamma = 8.4e-09.
\end{equation}

\begin{figure}[ht!]
    \centering
    \includegraphics[width=0.45\textwidth]{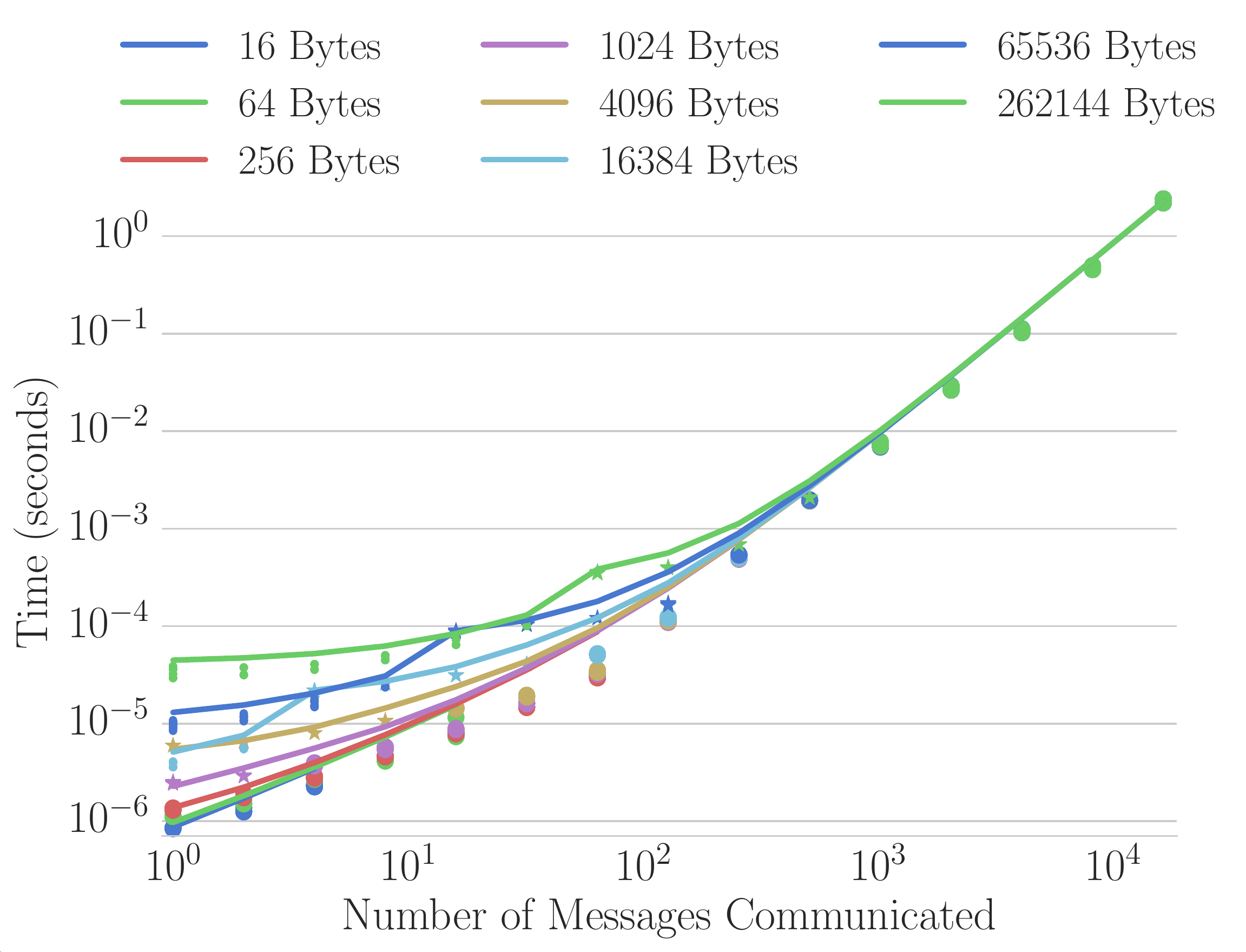}
    \caption{The measured versus modeled times for Blue Waters, where the model
    is a combination of the max-rate model and the contention, for
    \texttt{HighVolumePingPong} tests with a variety of message counts and sizes.  The
    receives are posted in the opposite order of which messages are received.}\label{figure:queue_search}
\end{figure}

Figure~\ref{figure:queue_search} shows the measured versus modeled times for the
\texttt{HighVolumePingPong} test in which the messages are received in the opposite order
from which they are posted.
This figure adds the queue search parameter to the original max-rate model,
yielding a more accurate analysis of the cost of large message counts.

\subsection{Network Contention}\label{section:network_contention}

\begin{figure}[ht!]
    \centering
    \includegraphics[width=0.45\textwidth]{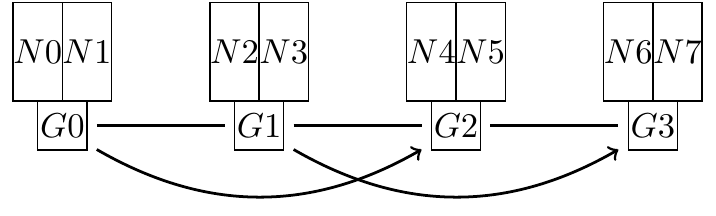}
    \caption{Four Gemini spanning a one-dimensional partition of the Blue Waters
      network.  A \texttt{HighVolumePingPong} test between all processes on Geminis $G0$
    and $G3$, and equivalent messages between $G1$ and $G3$ will result in a
    large amount of contention for the middle link in the partition.}\label{figure:gemini_line}
\end{figure}

When a large amount of data is communicated throughout the network, multiple
messages are often required to traverse the same link, yielding contention
within the network.  This network contention can occur on as few as eight nodes
of Blue Waters when a one-dimensional partition of the network is attained.
Figure~\ref{figure:gemini_line} shows a line of four Geminis, each containing
two nodes, with a one-dimension network partition.
Communicating messages from all $32$ processes on Gemini $0$ to corresponding
processes on Gemini $2$, and similarly sending from Geminis $1$ to $3$, requires
all data to traverse the network link between Geminis $1$ and $2$.  Therefore,
contention of this link occurs.

\begin{figure}[ht!]
    \centering
    \includegraphics[width=0.45\textwidth]{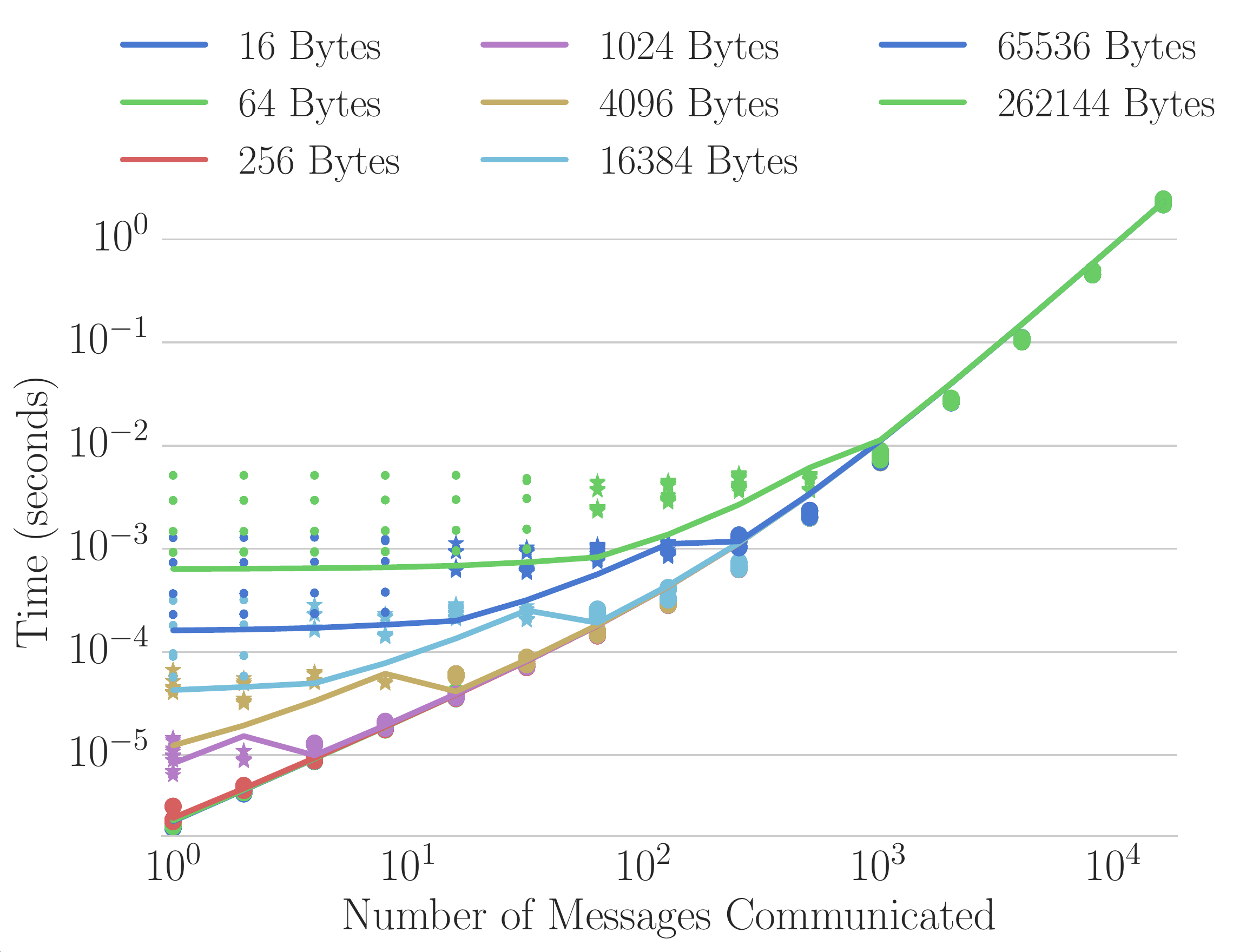}
    \caption{Measured versus modeled times for \texttt{HighVolumePingPong} communication
    among the sets of Blue Waters Geminis described in
    Figure~\ref{figure:gemini_line}.  The modeled times, which are a combination
    of max-rate models and the queue search parameters, do not capture the
    additional costs associated with contention.}\label{figure:contention_not_modeled}
\end{figure}

Figure~\ref{figure:contention_not_modeled} shows the measured and modeled costs
of sending messages of various counts and sizes among all processes on the
row of Geminis, where the model contains both the max-rate and queue search
measures.
The model underestimates the cost of communicating a large amount of data at
smaller message counts, before queue search time dominates.  The additional
measured cost can be modeled through an extra network contention parameter.
This addition measure is defined as
\begin{equation}
    T_{c} = \delta \cdot \ell,
\end{equation}
where $\delta$ is the per-byte penalty acquired waiting for a network link and
$\ell$ is the number of bytes to traverse each link.  Network contention only
occurs during inter-node communication.  However, the cost of contention is
constant regardless of the message sending protocol.  The measure for all
inter-node messages on Blue Waters is
\begin{equation}
    \delta = 1.0e-10.
\end{equation}

\begin{figure}[ht!]
    \centering
    \includegraphics[width=0.3\textwidth]{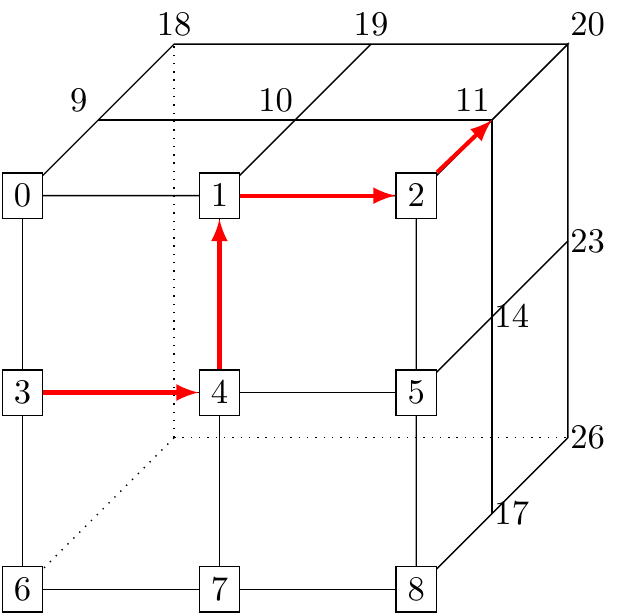}
    \caption{A perfect cube partition of Blue Waters Geminis is used to
    calculate the average number of hops traversed by each byte of data.  In
    this example, a message from a process on Gemini $3$ to Gemini $11$
    traverses $4$ network links.}\label{figure:cube}
\end{figure}

The number of bytes to traverse any link, or $\ell$, is dependent on the number
of links each message traverses.  Therefore, knowledge of the specific partition
of the network is required to model the associated cost.  This requirement is
removed by assuming the nodes are connected through a perfect three-dimension
cube portion of Blue Waters' three dimensional torus, as displayed in
Figure~\ref{figure:cube}.
Therefore, $\ell$ is defined as the following
\begin{equation}
    \ell = 2 h^{3} \cdot b \cdot \texttt{ppn},
    \label{equation:contention_links}
\end{equation}
\begin{figure*}[t] 
    \centering
    \includegraphics[width=0.45\textwidth]{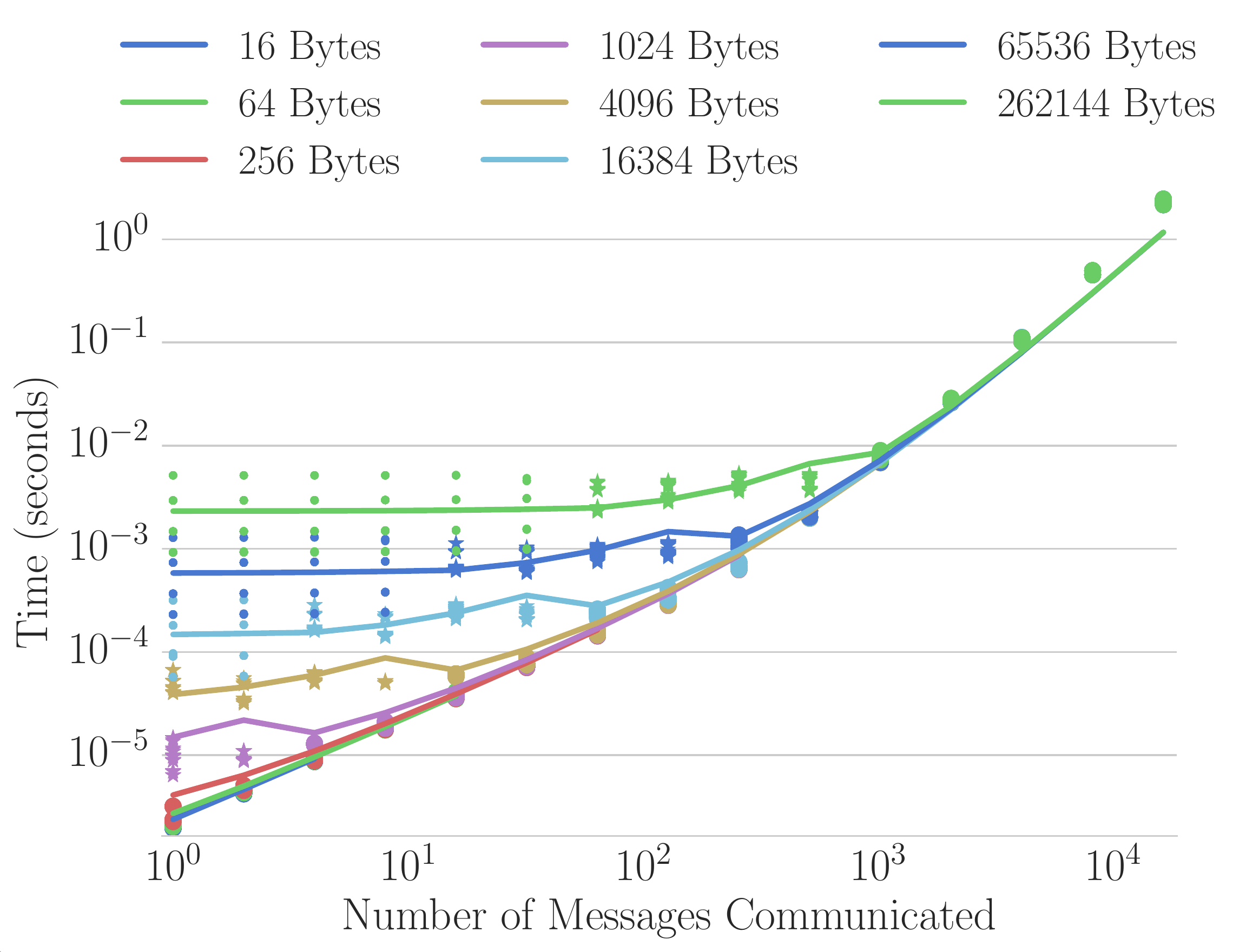}
    \caption{Modeled versus measured times for \texttt{HighVolumePingPong} communication
    about four Blue Waters Geminis
    spanning a one-dimensional partition of the networks.  The processes on the
    Geminis and nodes communicate as described in
    Figure~\ref{figure:gemini_line}.  The modeled times are a combination of
    max-rate models, queue search costs, and the contention parameter.}\label{figure:network_contention}
\end{figure*}
where $h$ is the average number of hops, or network links, traversed by each
byte of data, and $b$ is the average number of bytes to be sent from any
process.  Therefore, as $h^{3}$ yields the number of Geminis within $h$ hops of
a given link, this measure estimates network contention assuming all bytes that
can traverse one single link do.  Furthermore, $2 b \cdot \texttt{ppn}$
calculates the average number of bytes communicated from each Gemini.

Figure~\ref{figure:network_contention} displays the measured and modeled costs
of communicating a variety of message counts and sizes among four Geminis of
Blue Waters, with a one-dimensional network contention.  This figure includes
the max-rate, queue search, and network contention models, yielding improved
accuracy in the model.

\section{Applications}\label{section:applications}

Sparse matrix-vector (SpMV) and sparse matrix-matrix (SpGEMM) multiplication are
commonly used in a variety of applications such as numerical methods and graph
algorithms.  When matrices are sufficiently sparse, point-to-point communication
is used to send only necessary values to the processes that need them.

Algebraic multigrid (AMG) is a sparse linear solver comprised of matrix
operations.  An AMG hierarchy consists of successively coarser, but denser,
matrices.  Therefore, each level in the hierarchy decreases in dimension, but
often increases in the number of non-zeros per row.  The various levels require
a variety of communication patterns, as the finer levels require communication
of few large messages while coarse levels require communicating a larger number
of small messages.

This section focuses on modeling the cost of matrix operations throughout an AMG
hierarchy as communication costs vary drastically among the levels.  The
hierarchy is formed with classical AMG to solve a three-dimensional
unstructured linear elasticity problems formed with 
MFEM~\footnote{\url{http://mfem.org}}.
All tests are performed with
RAPtor~\cite{RAPtor} on $512$ nodes of Blue Waters. The original
linear elasticity system consists of $840\,000$ unknowns and $65$ million
non-zeros.

\begin{figure}[ht!]
    \centering
    \includegraphics[width=0.45\textwidth]{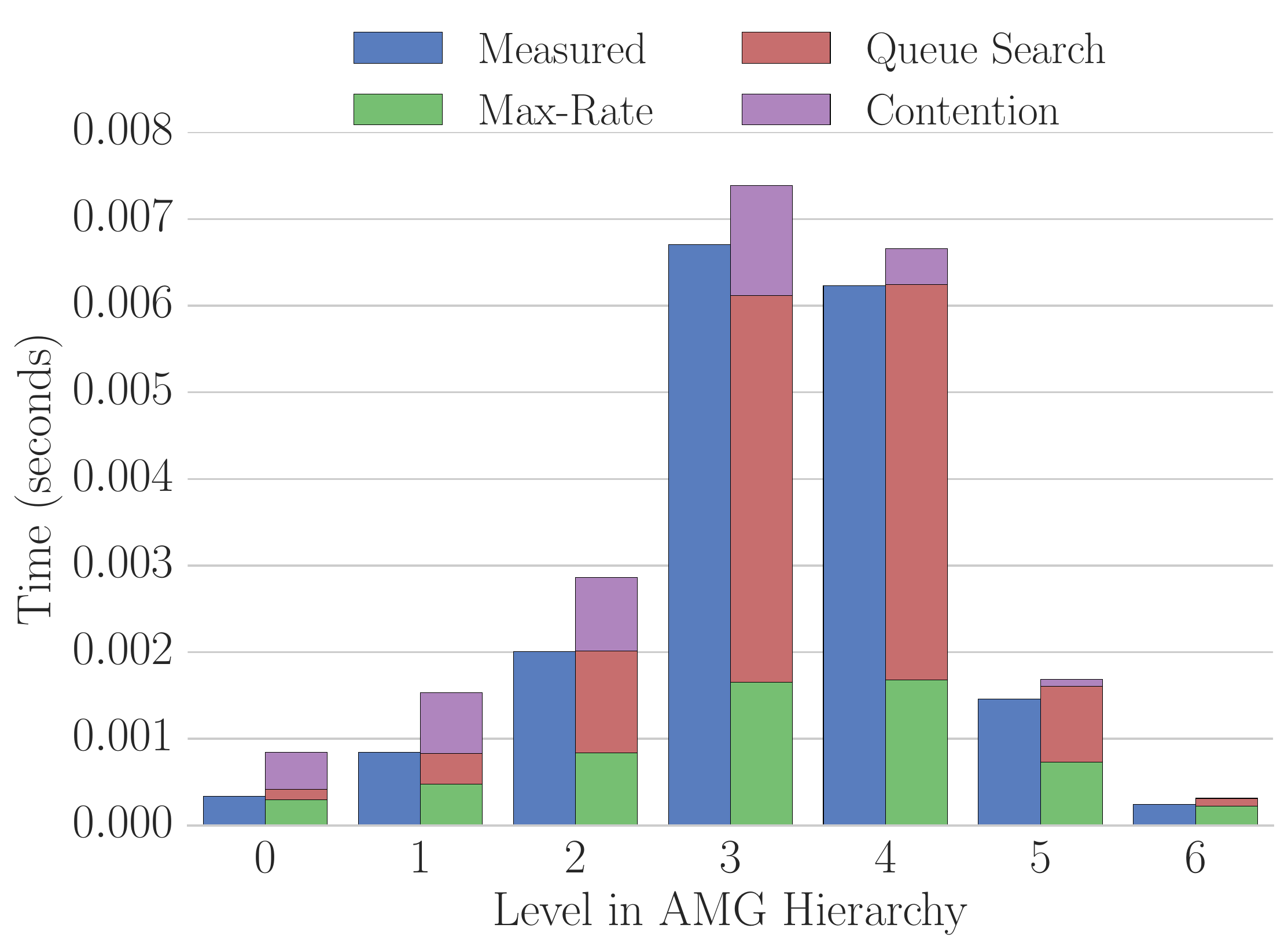}
    \caption{Measured versus modeled times for performing a \textbf{SpMV} on
    each level of linear elasticity AMG hierarchy.}\label{figure:model_spmv}
\end{figure}

Figure~\ref{figure:model_spmv} displays the measured and modeled costs for
performing a SpMV on each level of the AMG hierarchy.
The modeled costs are partitioned into the max-rate model, queue search costs,
and network contention penalties.  The cost of each SpMV is accurately captured
when all model parameters are included, with a large improvement over modeling
with only the max-rate model.  Furthermore, the model indicates that the
majority of communication costs on coarse levels near the middle of the
hierarchy are due mainly to queue search costs, motivating efforts for
minimizing the number of messages received or posted at any time.

\begin{figure}[ht!]
    \centering
    \includegraphics[width=0.45\textwidth]{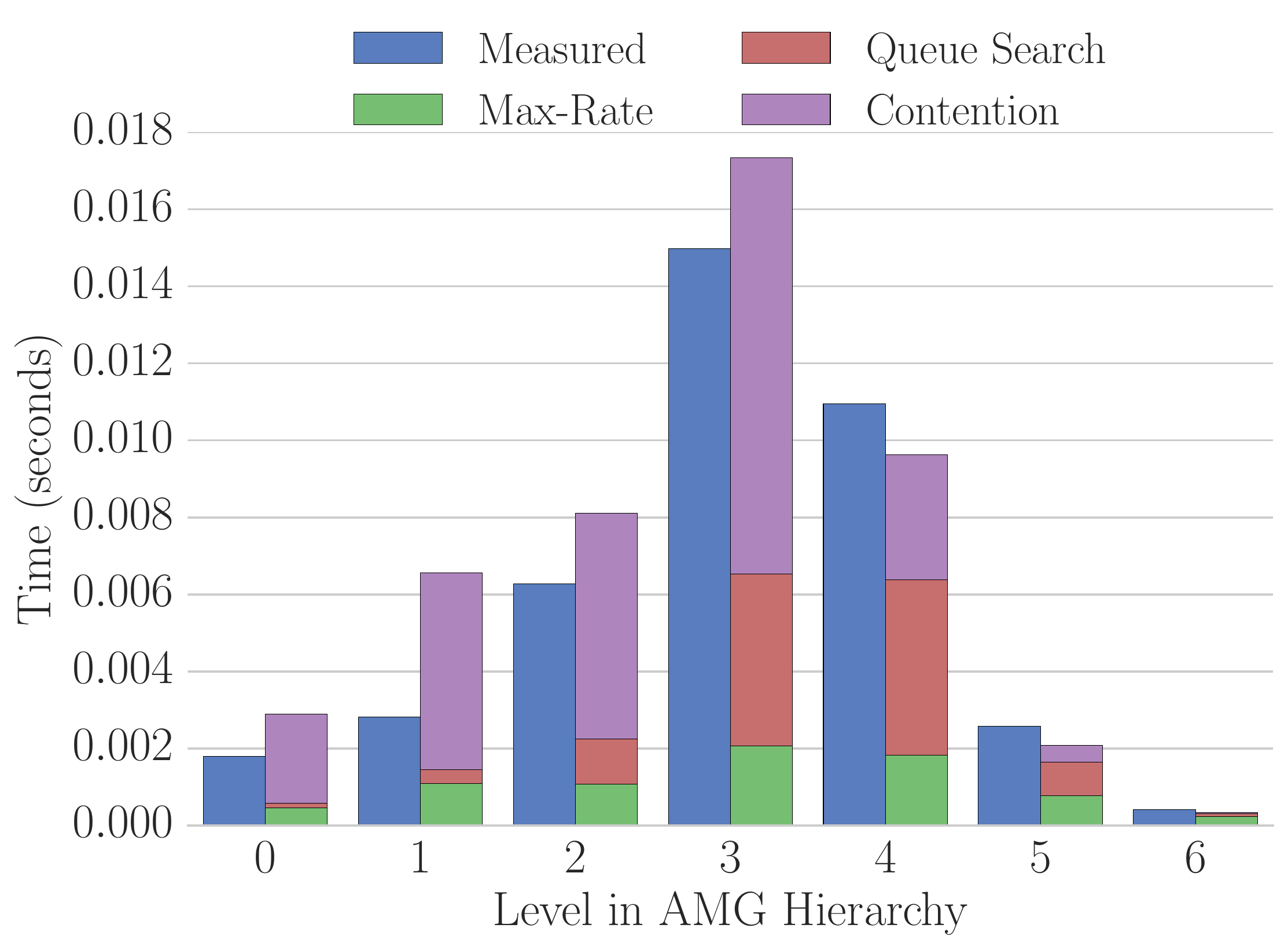}
    \caption{Measured versus modeled times for performing a \textbf{SpGEMM} on
    each level of linear elasticity AMG hierarchy.}\label{figure:model_spgemm}
\end{figure}

The measured and modeled costs for performing an SpGEMM on each level of the AMG
hierarchy are shown in Figure~\ref{figure:model_spgemm}.
The models are again partitioned into max-rate, queue search, and network
contention costs, displaying a large improvement in the model accuracy from the
combination of queue search and contention parameters.  These models show that
more significant costs of the SpGEMMs are from network contention, and while
queue search times could be reduced, larger savings would be acquired by
reducing the number of bytes to traverse any link.

Combining the max-rate model with queue search and network contention parameters
improves the accuracy of the model, but also over-predicts the timings.  This
over-prediction is a result of using an upper bound on the queue search
parameter, corresponding to the cost of posting receives in the opposite order
from which messages arrive.  The upper bound assumes the $\frac{n \cdot
(n+1)}{2}$ elements are searched, while only $n$ elements are searched if
the receives are posted in the correct order.  The queue search cost was
measured for each of these applications by probing for the first available
message and computing its position in the queue.  The maximum cost on any
process was consistently around $\frac{n \cdot n}{3}$, which is approximated
much more accurately by the upper bound than the lower.  Furthermore, reversing
the ordering of the receives results in a different process incurring the
maximum queue search penalty, but this cost stays constant.

\section{Conclusion}\label{section:conclusion}

Common applications of point-to-point communication typically require a large
number of messages to be communication, with large variability in corresponding
messages sizes.  Furthermore, there are often combinations of intra-socket,
intra-node, and inter-node messages, with the latter commonly traversing
multiple links of the network.  Therefore, the traditional postal and max-rate
models can be improved by splitting standard parameters into on-socket, on-node,
and off-node.  Additional parameters for bounding the queue search cost and
estimating network contention further improve the accuracy of these models.
When all parameters are used, the models accurately capture the cost of
irregular sparse matrix operations on Blue Waters.

The model parameters are all computed with ping-pong and \texttt{HighVolumePingPong}
tests on few nodes, with the majority of tests requiring only a single node
while network contention parameters are calculated on up to eight nodes.
However, these parameters remain accurate when modeling sparse matrix
operations on $512$ nodes, indicating this model can be extended to large core
counts with no additional work.

This model may need alterations to accurately capture the costs on different
systems.  For example, architectures with only a single socket per node, such as
Blue Gene Q machines, will only need to partition the max-rate model into
on-node and off-node messages.  Furthermore, MPI implementations with an
optimized queue search will require alternative queue search penalties, while
implementations with dynamic message routing will require a block of
communicating nodes to capture network contention in the test in Figure~\ref{figure:gemini_line}.

Limitations for this model include using the upper bound for queue search time
and assuming the process domain is mapped to a cube for network contention.  The
queue search time will overestimate the actual cost, while the accuracy of the
network contention penalty can vary with actual partition acquired.

These models can be further extended to include topology-aware parameters, such
as additional latency required for messages traversing a large number of links.
Furthermore, the models motivate future directions for tested applications, such
as methods for reducing queue search time in SpMVs and network contention in
SpGEMMs.

\begin{acks}
    This research is part of the Blue Waters sustained-petascale computing
    project, which is supported by the \grantsponsor{1234}{National Science
    Foundation}{https://www.nsf.gov/} (awards \grantnum{1234}{OCI-0725070} and
    \grantnum{1234}{ACI-1238993}) and the state of Illinois. Blue Waters is a
    joint effort of the University of Illinois at Urbana-Champaign and its
    National Center for Supercomputing Applications.  This material is based in
    part upon work supported by the \grantsponsor{4567}{National Science
    Foundation Graduate Research Fellowship
    Program}{https://www.nsf.gov/funding/pgm_summ.jsp?pims_id=6201} under Grant
    Number \grantnum{4567}{DGE-1144245}.  This material is based in part upon
    work supported by the \grantsponsor{5432}{Department of Energy, National
    Nuclear}{https://www.energy.gov/nnsa/national-nuclear-security-administration
    Security Administration}, under Award Number
    \grantnum{5432}{DE-NA0002374}. 
\end{acks}

\bibliographystyle{ACM-Reference-Format}
\bibliography{paper}


\begin{thebibliography}{17}


\ifx \showCODEN    \undefined \def \showCODEN     #1{\unskip}     \fi
\ifx \showDOI      \undefined \def \showDOI       #1{#1}\fi
\ifx \showISBNx    \undefined \def \showISBNx     #1{\unskip}     \fi
\ifx \showISBNxiii \undefined \def \showISBNxiii  #1{\unskip}     \fi
\ifx \showISSN     \undefined \def \showISSN      #1{\unskip}     \fi
\ifx \showLCCN     \undefined \def \showLCCN      #1{\unskip}     \fi
\ifx \shownote     \undefined \def \shownote      #1{#1}          \fi
\ifx \showarticletitle \undefined \def \showarticletitle #1{#1}   \fi
\ifx \showURL      \undefined \def \showURL       {\relax}        \fi
\providecommand\bibfield[2]{#2}
\providecommand\bibinfo[2]{#2}
\providecommand\natexlab[1]{#1}
\providecommand\showeprint[2][]{arXiv:#2}

\bibitem[\protect\citeauthoryear{Agarwal, Sharma, and Kal{\'e}}{Agarwal
  et~al\mbox{.}}{2006}]%
        {HopModel}
\bibfield{author}{\bibinfo{person}{Tarun Agarwal}, \bibinfo{person}{Amit
  Sharma}, {and} \bibinfo{person}{Laxmikant~V. Kal{\'e}}.}
  \bibinfo{year}{2006}\natexlab{}.
\newblock \showarticletitle{Topology-aware Task Mapping for Reducing
  Communication Contention on Large Parallel Machines}. In
  \bibinfo{booktitle}{\emph{Proceedings of the 20th International Conference on
  Parallel and Distributed Processing}} \emph{(\bibinfo{series}{IPDPS'06})}.
  \bibinfo{publisher}{IEEE Computer Society}, \bibinfo{address}{Washington, DC,
  USA}, \bibinfo{pages}{145--145}.
\newblock
\showISBNx{1-4244-0054-6}
\urldef\tempurl%
\url{http://dl.acm.org/citation.cfm?id=1898953.1899075}
\showURL{%
\tempurl}


\bibitem[\protect\citeauthoryear{Alexandrov, Ionescu, Schauser, and
  Scheiman}{Alexandrov et~al\mbox{.}}{1995}]%
        {LogGP}
\bibfield{author}{\bibinfo{person}{Albert Alexandrov},
  \bibinfo{person}{Mihai~F. Ionescu}, \bibinfo{person}{Klaus~E. Schauser},
  {and} \bibinfo{person}{Chris Scheiman}.} \bibinfo{year}{1995}\natexlab{}.
\newblock \showarticletitle{LogGP: Incorporating Long Messages into the LogP
  Model -- One Step Closer Towards a Realistic Model for Parallel Computation}.
  In \bibinfo{booktitle}{\emph{Proceedings of the Seventh Annual ACM Symposium
  on Parallel Algorithms and Architectures}} \emph{(\bibinfo{series}{SPAA
  '95})}. \bibinfo{publisher}{ACM}, \bibinfo{address}{New York, NY, USA},
  \bibinfo{pages}{95--105}.
\newblock
\showISBNx{0-89791-717-0}
\urldef\tempurl%
\url{https://doi.org/10.1145/215399.215427}
\showDOI{\tempurl}


\bibitem[\protect\citeauthoryear{Bienz and Olson}{Bienz and Olson}{2017}]%
        {RAPtor}
\bibfield{author}{\bibinfo{person}{Amanda Bienz} {and} \bibinfo{person}{Luke~N.
  Olson}.} \bibinfo{year}{2017}\natexlab{}.
\newblock \bibinfo{title}{{RAPtor}: parallel algebraic multigrid v0.1}.
\newblock
\newblock
\urldef\tempurl%
\url{https://github.com/lukeolson/raptor}
\showURL{%
\tempurl}
\newblock
\shownote{Release 0.1.}


\bibitem[\protect\citeauthoryear{Bode, Butler, Dunning, Hoefler, Kramer, Gropp,
  and Hwu}{Bode et~al\mbox{.}}{2013}]%
        {BlueWaters13}
\bibfield{author}{\bibinfo{person}{Brett Bode}, \bibinfo{person}{Michelle
  Butler}, \bibinfo{person}{Thom Dunning}, \bibinfo{person}{Torsten Hoefler},
  \bibinfo{person}{William Kramer}, \bibinfo{person}{William Gropp}, {and}
  \bibinfo{person}{Wen-mei Hwu}.} \bibinfo{year}{2013}\natexlab{}.
\newblock \showarticletitle{The {B}lue {W}aters Super-System for
  Super-Science}.
\newblock In \bibinfo{booktitle}{\emph{Contemporary High Performance
  Computing}}. \bibinfo{publisher}{Chapman and Hall/CRC},
  \bibinfo{pages}{339--366}.
\newblock
\showISBNx{978-1-4665-6834-1}
\urldef\tempurl%
\url{https://doi.org/10.1201/b14677-16}
\showDOI{\tempurl}


\bibitem[\protect\citeauthoryear{Cownie and Gropp}{Cownie and Gropp}{1999}]%
        {MPIQueues}
\bibfield{author}{\bibinfo{person}{James Cownie} {and} \bibinfo{person}{William
  Gropp}.} \bibinfo{year}{1999}\natexlab{}.
\newblock \showarticletitle{A Standard Interface for Debugger Access to Message
  Queue Information in MPI}. In \bibinfo{booktitle}{\emph{Recent Advances in
  Parallel Virtual Machine and Message Passing Interface}},
  \bibfield{editor}{\bibinfo{person}{Jack Dongarra}, \bibinfo{person}{Emilio
  Luque}, {and} \bibinfo{person}{Tom{\`a}s Margalef}} (Eds.).
  \bibinfo{publisher}{Springer Berlin Heidelberg}, \bibinfo{address}{Berlin,
  Heidelberg}, \bibinfo{pages}{51--58}.
\newblock


\bibitem[\protect\citeauthoryear{Culler, Karp, Patterson, Sahay, Santos,
  Schauser, Subramonian, and von Eicken}{Culler et~al\mbox{.}}{1996}]%
        {LogP}
\bibfield{author}{\bibinfo{person}{David~E. Culler},
  \bibinfo{person}{Richard~M. Karp}, \bibinfo{person}{David Patterson},
  \bibinfo{person}{Abhijit Sahay}, \bibinfo{person}{Eunice~E. Santos},
  \bibinfo{person}{Klaus~Erik Schauser}, \bibinfo{person}{Ramesh Subramonian},
  {and} \bibinfo{person}{Thorsten von Eicken}.}
  \bibinfo{year}{1996}\natexlab{}.
\newblock \showarticletitle{LogP: A Practical Model of Parallel Computation}.
\newblock \bibinfo{journal}{\emph{Commun. ACM}} \bibinfo{volume}{39},
  \bibinfo{number}{11} (\bibinfo{date}{Nov.} \bibinfo{year}{1996}),
  \bibinfo{pages}{78--85}.
\newblock
\showISSN{0001-0782}
\urldef\tempurl%
\url{https://doi.org/10.1145/240455.240477}
\showDOI{\tempurl}


\bibitem[\protect\citeauthoryear{D\'{o}zsa, Kumar, Balaji, Buntinas, Goodell,
  Gropp, Ratterman, and Thakur}{D\'{o}zsa et~al\mbox{.}}{2010}]%
        {QueueSearch}
\bibfield{author}{\bibinfo{person}{G\'{a}bor D\'{o}zsa},
  \bibinfo{person}{Sameer Kumar}, \bibinfo{person}{Pavan Balaji},
  \bibinfo{person}{Darius Buntinas}, \bibinfo{person}{David Goodell},
  \bibinfo{person}{William Gropp}, \bibinfo{person}{Joe Ratterman}, {and}
  \bibinfo{person}{Rajeev Thakur}.} \bibinfo{year}{2010}\natexlab{}.
\newblock \showarticletitle{Enabling Concurrent Multithreaded MPI Communication
  on Multicore Petascale Systems}. In \bibinfo{booktitle}{\emph{Proceedings of
  the 17th European MPI Users' Group Meeting Conference on Recent Advances in
  the Message Passing Interface}} \emph{(\bibinfo{series}{EuroMPI'10})}.
  \bibinfo{publisher}{Springer-Verlag}, \bibinfo{address}{Berlin, Heidelberg},
  \bibinfo{pages}{11--20}.
\newblock
\showISBNx{3-642-15645-2, 978-3-642-15645-8}
\urldef\tempurl%
\url{http://dl.acm.org/citation.cfm?id=1894122.1894125}
\showURL{%
\tempurl}


\bibitem[\protect\citeauthoryear{Flajslik, Dinan, and Underwood}{Flajslik
  et~al\mbox{.}}{2016}]%
        {MultiQueueSearch}
\bibfield{author}{\bibinfo{person}{Mario Flajslik}, \bibinfo{person}{James
  Dinan}, {and} \bibinfo{person}{Keith~D. Underwood}.}
  \bibinfo{year}{2016}\natexlab{}.
\newblock \showarticletitle{Mitigating MPI Message Matching Misery}. In
  \bibinfo{booktitle}{\emph{High Performance Computing}},
  \bibfield{editor}{\bibinfo{person}{Julian~M. Kunkel}, \bibinfo{person}{Pavan
  Balaji}, {and} \bibinfo{person}{Jack Dongarra}} (Eds.).
  \bibinfo{publisher}{Springer International Publishing},
  \bibinfo{address}{Cham}, \bibinfo{pages}{281--299}.
\newblock


\bibitem[\protect\citeauthoryear{Frank, Agarwal, and Vernon}{Frank
  et~al\mbox{.}}{1997}]%
        {LoPC}
\bibfield{author}{\bibinfo{person}{Matthew~I. Frank}, \bibinfo{person}{Anant
  Agarwal}, {and} \bibinfo{person}{Mary~K. Vernon}.}
  \bibinfo{year}{1997}\natexlab{}.
\newblock \showarticletitle{{LoPC}: Modeling Contention in Parallel
  Algorithms}. In \bibinfo{booktitle}{\emph{Proceedings of the Sixth ACM
  SIGPLAN Symposium on Principles and Practice of Parallel Programming}}
  \emph{(\bibinfo{series}{PPOPP '97})}. \bibinfo{publisher}{ACM},
  \bibinfo{address}{New York, NY, USA}, \bibinfo{pages}{276--287}.
\newblock
\showISBNx{0-89791-906-8}
\urldef\tempurl%
\url{https://doi.org/10.1145/263764.263803}
\showDOI{\tempurl}


\bibitem[\protect\citeauthoryear{Gibbons}{Gibbons}{1989}]%
        {PRAM}
\bibfield{author}{\bibinfo{person}{P.~B. Gibbons}.}
  \bibinfo{year}{1989}\natexlab{}.
\newblock \showarticletitle{A More Practical PRAM Model}. In
  \bibinfo{booktitle}{\emph{Proceedings of the First Annual ACM Symposium on
  Parallel Algorithms and Architectures}} \emph{(\bibinfo{series}{SPAA '89})}.
  \bibinfo{publisher}{ACM}, \bibinfo{address}{New York, NY, USA},
  \bibinfo{pages}{158--168}.
\newblock
\showISBNx{0-89791-323-X}
\urldef\tempurl%
\url{https://doi.org/10.1145/72935.72953}
\showDOI{\tempurl}


\bibitem[\protect\citeauthoryear{Gropp, Olson, and Samfass}{Gropp
  et~al\mbox{.}}{2016}]%
        {MaxRate}
\bibfield{author}{\bibinfo{person}{William Gropp}, \bibinfo{person}{Luke~N.
  Olson}, {and} \bibinfo{person}{Philipp Samfass}.}
  \bibinfo{year}{2016}\natexlab{}.
\newblock \showarticletitle{Modeling {MPI} Communication Performance on {SMP}
  Nodes: Is It Time to Retire the Ping Pong Test}. In
  \bibinfo{booktitle}{\emph{Proceedings of the 23rd European MPI Users' Group
  Meeting}} \emph{(\bibinfo{series}{EuroMPI 2016})}. \bibinfo{publisher}{ACM},
  \bibinfo{address}{New York, NY, USA}, \bibinfo{pages}{41--50}.
\newblock
\urldef\tempurl%
\url{https://doi.org/10.1145/2966884.2966919}
\showDOI{\tempurl}


\bibitem[\protect\citeauthoryear{Hoefler, Schneider, and Lumsdaine}{Hoefler
  et~al\mbox{.}}{2010}]%
        {LogGOPSim}
\bibfield{author}{\bibinfo{person}{T. Hoefler}, \bibinfo{person}{T. Schneider},
  {and} \bibinfo{person}{A. Lumsdaine}.} \bibinfo{year}{2010}\natexlab{}.
\newblock \showarticletitle{{LogGOPSim - Simulating Large-Scale Applications in
  the LogGOPS Model}}. In \bibinfo{booktitle}{\emph{Proceedings of the 19th ACM
  International Symposium on High Performance Distributed Computing}}.
  \bibinfo{publisher}{ACM}, \bibinfo{address}{Chicago, Illinois},
  \bibinfo{pages}{597--604}.
\newblock
\showISBNx{978-1-60558-942-8}


\bibitem[\protect\citeauthoryear{Jain, Bhatele, Robson, Gamblin, and Kale}{Jain
  et~al\mbox{.}}{2013}]%
        {PredictContention}
\bibfield{author}{\bibinfo{person}{Nikhil Jain}, \bibinfo{person}{Abhinav
  Bhatele}, \bibinfo{person}{Michael~P. Robson}, \bibinfo{person}{Todd
  Gamblin}, {and} \bibinfo{person}{Laxmikant~V. Kale}.}
  \bibinfo{year}{2013}\natexlab{}.
\newblock \showarticletitle{Predicting Application Performance Using Supervised
  Learning on Communication Features}. In \bibinfo{booktitle}{\emph{Proceedings
  of the International Conference on High Performance Computing, Networking,
  Storage and Analysis}} \emph{(\bibinfo{series}{SC '13})}.
  \bibinfo{publisher}{ACM}, \bibinfo{address}{New York, NY, USA}, Article
  \bibinfo{articleno}{95}, \bibinfo{numpages}{12}~pages.
\newblock
\showISBNx{978-1-4503-2378-9}
\urldef\tempurl%
\url{https://doi.org/10.1145/2503210.2503263}
\showDOI{\tempurl}


\bibitem[\protect\citeauthoryear{Moritz and Frank}{Moritz and Frank}{1998}]%
        {LoGPC}
\bibfield{author}{\bibinfo{person}{Csaba~Andras Moritz} {and}
  \bibinfo{person}{Matthew~I. Frank}.} \bibinfo{year}{1998}\natexlab{}.
\newblock \showarticletitle{{LoGPC}: Modeling Network Contention in
  Message-passing Programs}. In \bibinfo{booktitle}{\emph{Proceedings of the
  1998 ACM SIGMETRICS Joint International Conference on Measurement and
  Modeling of Computer Systems}} \emph{(\bibinfo{series}{SIGMETRICS
  '98/PERFORMANCE '98})}. \bibinfo{publisher}{ACM}, \bibinfo{address}{New York,
  NY, USA}, \bibinfo{pages}{254--263}.
\newblock
\showISBNx{0-89791-982-3}
\urldef\tempurl%
\url{https://doi.org/10.1145/277851.277933}
\showDOI{\tempurl}


\bibitem[\protect\citeauthoryear{Saboo, Singla, Unger, and Kale}{Saboo
  et~al\mbox{.}}{2001}]%
        {Bigsim}
\bibfield{author}{\bibinfo{person}{N. Saboo}, \bibinfo{person}{A.~K. Singla},
  \bibinfo{person}{J.~M. Unger}, {and} \bibinfo{person}{L.~V. Kale}.}
  \bibinfo{year}{2001}\natexlab{}.
\newblock \showarticletitle{Emulating petaflops machines and blue gene}. In
  \bibinfo{booktitle}{\emph{Proceedings 15th International Parallel and
  Distributed Processing Symposium. IPDPS 2001}}. \bibinfo{publisher}{IEEE},
  \bibinfo{address}{San Francisco, CA, USA}, \bibinfo{pages}{2084--2091}.
\newblock
\showISSN{1530-2075}
\urldef\tempurl%
\url{https://doi.org/10.1109/IPDPS.2001.925206}
\showDOI{\tempurl}


\bibitem[\protect\citeauthoryear{Schneider, Hoefler, and Lumsdaine}{Schneider
  et~al\mbox{.}}{2009}]%
        {ORCS}
\bibfield{author}{\bibinfo{person}{T. Schneider}, \bibinfo{person}{T. Hoefler},
  {and} \bibinfo{person}{A. Lumsdaine}.} \bibinfo{year}{2009}\natexlab{}.
\newblock \bibinfo{booktitle}{\emph{{ORCS: An Oblivious Routing Congestion
  Simulator}}}.
\newblock \bibinfo{type}{{T}echnical {R}eport} 675.
  \bibinfo{institution}{Indiana University}.
\newblock


\bibitem[\protect\citeauthoryear{Steffenel}{Steffenel}{2006}]%
        {ContentionAlltoall}
\bibfield{author}{\bibinfo{person}{L.~A. Steffenel}.}
  \bibinfo{year}{2006}\natexlab{}.
\newblock \showarticletitle{Modeling Network Contention Effects on All-to-All
  Operations}. In \bibinfo{booktitle}{\emph{2006 IEEE International Conference
  on Cluster Computing}}. \bibinfo{publisher}{IEEE},
  \bibinfo{address}{Barcelona, Spain}, \bibinfo{pages}{1--10}.
\newblock
\showISSN{1552-5244}
\urldef\tempurl%
\url{https://doi.org/10.1109/CLUSTR.2006.311889}
\showDOI{\tempurl}


\end{thebibliography}

\end{document}